\documentclass[12pt,dvips]{article}
\usepackage{epsfig}

\begin{document}

\title{\bf Newtonian Quantum Gravity}

\author {Johan Hansson\footnote{c.johan.hansson@ltu.se} \\
 Department of Physics, Lule{\aa} University of Technology
 \\ SE-971 87 Lule\aa, Sweden}

\date{}

\maketitle

\begin{abstract}
%\textit{Prelimin\"{a}rt...}
A Newtonian approach to quantum gravity is studied. At least for
weak gravitational fields it should be a valid approximation. Such
an approach could be used to point out problems and prospects
inherent in a more exact theory of quantum gravity, yet to be
discovered. Newtonian quantum gravity, e.g., shows promise for
prohibiting black holes altogether (which would eliminate
singularities and also solve the black hole information paradox),
breaks the equivalence principle of general relativity, and
supports non-local interactions (quantum entanglement). Its
predictions should also be testable at length scales well above
the ``Planck scale", by high-precision experiments feasible even
with existing technology. As an illustration of the theory, it
turns out that the solar system, superficially, perfectly well can
be described as a quantum gravitational system, provided that the
$l$ quantum number has its maximum value, $n-1$. This results
exactly in Kepler's third law. If also the $m$ quantum number has
its maximum value ($\pm l$) the probability density has a very
narrow torus-like form, centered around the classical planetary
orbits. However, as the probability density is independent of the
azimuthal angle $\phi$ there is, from quantum gravity arguments,
no reason for planets to be located in any unique place along the
orbit (or even \textit{in} an orbit for $m \neq \pm l$). This is,
in essence, a reflection of the ``measurement problem" inherent in
all quantum descriptions.
\\
\\
%PACS numbers: 03.65.-w, 03.65.Bz
\end{abstract}

%%\pacs{ PACS numbers: 03.65.-w, 03.65.Bz 11.15.-q }
%%\newpage

The greatest fundamental challenge facing theoretical physics has
for many years been to reconcile gravity with quantum physics.
There have been numerous attempts to do so, but so far there is no
established and experimentally/observationally tested theory of
``quantum gravity", the two main contenders presently being string
theory \cite{strings} and loop quantum gravity \cite{loop}, with
``outsiders" like twistor theory \cite{twistors}, non-commutative
geometry \cite{connes}, etc.

The motivations for studying newtonian quantum gravity are:

1) Quantum theory is supposed to be universal, i.e., it should be
valid on all length scales and for all objects, as there in
principle exists no size/charge/mass-limit to its applicability.
In atomic physics the practical restriction comes about due to the
fact that there is a limit to arbitrarily large atomic nuclei as,
i) the Coulomb force between protons is repulsive, eventually
overpowering the strong nuclear force trying to hold the nucleus
together, ii) the additional weak force makes neutron-rich nuclei
decay before they grow too large. Also, the electric charge comes
in both positive and negative, and as a result a big lump of
matter is almost always electrically neutral\footnote{The same
also applies for e.g. the strong force, as the three different
color charges (``red", ``green", ``blue") always combine to
produce color-neutral hadrons and bulk matter.}. Neither of these
effects are present in ``pure" quantum gravity.

2) For weak gravitational fields the newtonian theory should be
sufficient. The weak-field newtonian limit is even used for
determining the constant $\kappa$ in Einstein's field equations of
general relativity $G_{\mu \nu} = \kappa T_{\mu \nu}$. The
newtonian limit is also almost always sufficient for practical
purposes in non-quantum gravity, except for a handful of extreme
cases (notably black holes and the very early universe), although
high-precision experiments in e.g. the solar system can and do
show deviations from the newtonian theory, always in favor of
general relativity \cite{Will}.

3) Even for strong gravitational fields the newtonian picture
gives the same prediction as general relativity for the
Schwarzschild radius of a spherically symmetric, non-rotating
black hole, and correct order of magnitude results for neutron
stars and cosmology. This could make it possible to deduce at
least qualitative results about strongly coupled quantum gravity,
as the newtonian viewpoint should give reliable first order
quantum gravitational results.

On the other hand would any ``absurd" results obtained from
newtonian quantum gravity, deviating from observations, implicate
either that:

A) General relativity cannot be quantized\footnote{This is an
automatic consequence of ``emergent" gravity, e.g. Sakharov's
theory \cite{Sakharov}, where gravity is a non-fundamental
interaction and rather a macroscopic consequence of other forces
and fields.}. An unsuccessful special case (the weak field limit)
would disprove the general case, whereas the opposite is not true.
(A vindicated weak field limit will not prove that the general
theory is also correct.)

or

B) Quantum mechanics fails at ``macroscopic" distances and for
macroscopic objects. This would mean that we in gravity have a
unique opportunity to understand the ``measurement problem" in
quantum mechanics, as proposed by e.g. K\'{a}rolyh\'{a}zy
\cite{Karolyhazy} and Penrose \cite{Penrose}. In that case we can
use gravity to probe the transition between quantum $\rightarrow$
classical behavior in detail, i.e. get experimental facts on
where, how and when the inherently undecided, subjective quantum
world of superpositions turns into the familiar objective
classical everyday world around us. One could, at least in
principle, envisage a test carried out in a free-falling (e.g.
satellite) environment where one alters $m$ (the gravitational
``test-charge") and $M$ (the gravitational ``source-charge") until
the expected quantum gravity results are observed, to obtain a
limit of where the quantum mechanical treatment breaks down, hance
making an experimental determination of the border between
``quantum" and ``classical", i.e. solving the quantum mechanical
measurement problem.
%It is indeed my own conviction that fundamental quantum gravity
%and the quantum mechanical measurement problem are intertwined and
%must be solved simultaneously in any successful approach.

In newtonian quantum gravity, at least as long as the system can
be approximately treated as a 2-body problem, it is possible to
use the mathematical identity between the electrostatic Coulomb
force in the hydrogen atom, and Newton's static gravitational
force under the substitution $e^2 / 4 \pi \epsilon_0 \rightarrow
GmM$. Therefore all analytical results from elementary quantum
physics directly lifts over to the quantum gravity case. For weak
electromagnetic fields, as in the hydrogen atom, the
electrodynamic corrections to the static Coulomb field are very
small, making the approximation excellent. The same applies to
quantum gravity, dynamical effects from general relativity are
negligible to a very high degree for weak gravitational fields. A
gravitationally bound 2-body system should then exhibit exactly
the same type of ``spectrum" as a hydrogen atom, but emitted in
(unobservable) graviton form instead of photons (easily detectable
as atomic spectra already in the 19th century).

For a free-falling 2-body system, e.g. in a satellite experiment
enclosed in a spherical vessel, it should in principle be possible
to measure the excitation energies for a suitable system. An
analogous result has seemingly already been accomplished for
neutrons in the gravitational field of the earth
\cite{Nezvishevsky}, although there are some quantum gravity
ambiguities as noted below.

For hydrogen-like (one electron) atoms, in the dominant Coulomb
central-field approximation, the energy levels depend only on the
principal quantum number, $n = 1, 2, 3, ...$
\begin{equation}
E_n = - \frac{m e^4}{8 \epsilon_0^2 h^2}\frac{Z^2}{n^2} = - E_H
\frac{Z^2}{n^2},
\end{equation}
where $E_H \simeq 13.6$ eV is the ionization energy, i.e. the
energy required to free the electron from the proton, and $Z$ the
number of protons in the nucleus.

The Bohr-radius, $a_0$, the innermost radius of circular orbits in
the old semi-classical Bohr-model, and also the distance $r$ for
which the probability density of the Schr\"{o}dinger equation for
the Hydrogen ground-state peaks, is
\begin{equation}
a_0 = \frac{h^2 \epsilon_0}{\pi m e^2},
\end{equation}
whereas the expectation value for the electron-nucleus separation
is
\begin{equation}
\langle r \rangle_{hydrogen}  \simeq \frac{n^2}{Z} a_0 = \frac{n^2
h^2 \epsilon_0}{Z \pi m e^2}.
\end{equation}
A comparison between the Coulomb potential in Hydrogen-like atoms
\begin{equation}
V_{hydrogen} = - \frac{Z e^2}{4 \pi \epsilon_0 r},
\end{equation}
and the Newtonian gravitational potential between two masses $m$
and $M$
\begin{equation}
V_{grav} = - \frac{G m M}{r},
\end{equation}
allows us to obtain all results of the gravitational case by the
simple substitution
\begin{equation}
\frac{Z e^2}{4 \pi \epsilon_0} \rightarrow G m M,
\end{equation}
in the well-known formulas for the Hydrogen atom.

For instance, the gravitational ``Bohr-radius", $b_0$, becomes
\begin{equation}
b_0 = \frac{h^2}{4 \pi^2 G m^2 M} = \frac{\hbar^2}{Gm^2M},
\end{equation}
and the quantum-gravitational energy levels are
\begin{equation}
E_n (grav) = - \frac{2 \pi^2 G^2 m^3 M^2}{h^2}\frac{1}{n^2} = -
\frac{G^2 m^3 M^2}{2 \hbar^2}\frac{1}{n^2} = - E_g \frac{1}{n^2},
\end{equation}
here again $E_g = G^2 m^3 M^2/ 2 \hbar^2$ is the energy required
to totally free the mass $m$ from $M$ in analogy to the Hydrogen
case, whereas the expectation value for the separation is
\begin{equation}
\langle r \rangle_{grav}  \simeq n^2 b_0 = \frac{n^2 \hbar^2}{G
m^2 M}.
\end{equation}
Also all the analytical solutions to the Schr\"{o}dinger equation,
the hydrogen wave-functions, carry over to the gravitational case
with the simple substitution $a_0 \rightarrow b_0$.
\begin{equation}
\psi_{nlm}  = R(r) \Theta(\theta) \Phi(\phi) = N_{nlm} R_{nl}
Y_{lm},
\end{equation}
where $N_{nlm}$ is the normalization constant, $R_{nl}$ the radial
wavefunction, and $Y_{lm}$, the spherical harmonics, contain the
angular part of the wavefunction.

Let us examine some concrete cases to obtain a feeling for these
relations: For a two-body problem composed of proton and electron
$b_0 \simeq 10^{29}$ m, several orders of magnitude larger than
the size of the observable universe ($\simeq 10^{26}$ m), whereas
$E_g \simeq 10^{-78}$ eV. For two neutrons $b_0 \simeq 10^{22}$ m,
$E_g \simeq 10^{-68}$ eV. For the earth and sun (approximated as a
two-body problem for illustrative reasons) one gets $b_0 \simeq
10^{-138}$ m, an absurdly small ground state separation, and $E_g
\simeq 10^{182}$ J, which is unphysical as the binding energy $E_g
\gg mc^2 \simeq 10^{42}$ J. We will see below how to deal with
these ``unphysical" cases and how the physical picture somewhat
surprisingly is connected to the Schwarzschild radius. For a
better 2-body application, let us consider a binary neutron star
system (one solar mass each), $b_0 \simeq 10^{-148}$ m, $E_g
\simeq 10^{198}$ J $\gg mc^2 \simeq 10^{47}$ J, again unphysical.
One could also ask how much $m$ would have to be in a
gravitational binary system (taking $m = M$) in order for $b_0$ to
be, for example, one meter: $m \simeq 10^{-19} kg$, or the mass of
a small virus. For a pair of ``Planck-objects" $m = M \simeq
10^{-8}$ kg, we get, maybe not surprisingly, $b_0 \simeq 10^{-35}$
m (the ``Planck length") and $E_g \simeq 10^{9}$ J (the ``Planck
energy") which also happens to be equal to $mc^2$. We could also
ask for the binary system mass (again taking $m=M$) giving exactly
the same numerical energy spectrum as for the Hydrogen atom, i.e.
taking $E_g = E_H = 13.6$ eV, resulting in $m \simeq 10^{-13}$ kg,
the mass of one human cell, and $b_0 \simeq 10^{-19}$ m.

One could ask for the mass, $m$, required to produce exactly the
quantum gravitational energy spectrum of hydrogen in a
gravitational field like that of earth, $M = M_{\oplus} \simeq 6
\times 10^{24}$ kg. This turns out to be $m \simeq 10^{-38}$ kg,
or an equivalent mass-energy of $\sim 10^{-3}$ eV, comparable to
the conjectured mass of neutrinos. As $b_0 \sim 1 \mu$m in this
case, only very highly excited states would be possible above the
earth surface. The matter would of course be quite different
around cosmic compact objects, for example the conjectured ``preon
stars" with masses comparable to the earth's and radii $\sim b_0$
\cite{HanssonSandin}.

We notice (e.g. through $b_0$) that the planets in the solar
system must be in very highly excited quantum gravitational
states. In that sense they are analogous to electrons in ``Rydberg
atoms" in atomic physics \cite{Rydberg}. To obtain a good two-body
approximation, let us study the sun-Jupiter system in a little
more detail.

For excited states with $l \neq 0$, and very large $n$ and $l$,
the expectation value of the distance is
\begin{equation}
\langle r \rangle \simeq \frac{1}{2}(3 n^2 - l^2) b_0,
\end{equation}
however as that is for an ensemble (average over many
measurements), for a single state it is in principle more
appropriate to use the most probable radial distance (``radius" of
orbital)
\begin{equation}
\tilde{r} = n^2 b_0,
\end{equation}
as a measure for the expected separation. However, for $n$ large
and $l=l_{max} = n-1$ the two coincide so that $\langle r \rangle
= \tilde{r}$

The angular momentum for Jupiter around the sun is $L \simeq 2
\times 10^{43}$ Js, giving an $l$-quantum number of $l = L/\hbar
\simeq 2 \times 10^{77}$. The most probable sun-Jupiter distance
is given by $\tilde{r} = n^2 b_0 \geq l^2 b_0 \simeq 7.6 \times
10^{11}$ m, which is the same as the actual separation. $E_n =
-E_g/n^2 \simeq - 1.6 \times 10^{35}$ J, so the magnitude of the
binding energy is much less than $mc^2 \simeq 1.8 \times 10^{44}$
J, making it physically allowed, and also of the same order of
magnitude as its classical counterpart $-GmM/r \simeq - 3.4 \times
10^{35}$ J. The sun-Jupiter system can thus seemingly be treated
as a quantum gravitational 2-body system, provided that it is
taken to have its maximally allowed value for its angular momentum
($l \simeq n$).

In fact, it is easy to show that for Kepler's law to apply, $l$
must be very close to $n$:

The period of revolution can be written

\begin{equation}
T = \frac{2 \pi m \tilde{r}^2}{L} = \frac{2 \pi m \tilde{r}^2}{l
\hbar},
\end{equation}
and assuming maximality for the angular momentum, $l \simeq n$,
gives

\begin{equation}
T  \simeq \frac{2 \pi m \tilde{r}^2}{n \hbar}.
\end{equation}
Solving the most probable distance, Eq. (12), for $n$ gives
\begin{equation}
n  = \frac{m \sqrt{GM\tilde{r}}}{\hbar},
\end{equation}
so that
\begin{equation}
T  \simeq \frac{2 \pi \tilde{r}^{3/2}}{\sqrt{GM}},
\end{equation}
which exactly is Kepler's law. So, the conclusion is that all the
planets in the solar system are in maximally allowed angular
momentum states quantum mechanically. The $l \simeq n$ quantum
numbers are as follows: $l_{sun} \simeq 2 \times 10^{75}$,
$l_{mercury} \simeq 8 \times 10^{72}$, $l_{venus} \simeq 2 \times
10^{74}$, $l_{earth} \simeq 3 \times 10^{74}$, $l_{mars} \simeq 4
\times 10^{73}$, $l_{jupiter} \simeq 2 \times 10^{77}$,
$l_{saturn} \simeq 8 \times 10^{76}$, $l_{uranus} \simeq 2 \times
10^{76}$, $l_{neptune} \simeq 2 \times 10^{76}$, $l_{pluto} \simeq
3 \times 10^{72}$. Even though the maximality of $L$ and $L_z$ are
automatic from the classical description, it is far from obvious
why the same should result from the more fundamental quantum
treatment, as noted below.

For states with $l = l_{max} = n-1$ and $m = \pm l$: i) There is
only one peak, at $r = \tilde{r}$, for the radial probability
density, and the ``spread" (variance) in the $r$-direction is
given by\footnote{The hydrogen wavefunctions for the gravitational
case give $\langle r^2 \rangle = [5n^2 + 1 -3l(l+1)] n^2 b_0^2 /2$
and $\langle r \rangle = [3n^2 - l(l+1)]b_0 /2$ .} $\Delta r =
\sqrt{\langle r^2 \rangle - \langle r \rangle^2} \simeq n^{3/2}
b_0 /2$. For the earth-sun system it means $\Delta r \simeq
10^{-26}$ m, ii) The angular $\theta$-part of the wavefunction for
maximal $m$-quantum number $|m| = l$, is $\propto \sin^{l}\theta$.
The probability density thus goes as $sin^{2l} \theta$ in the
$\theta$-direction, meaning that only $\theta = \pi /2$ is
nonvanishing for large $l$. The azimuthal ($\phi$) part of the
angular wavefunction $Y_{lm}$ is purely imaginary, making it drop
out of the probability density, so that \textit{all} values of
$\phi$ are equally likely. (This $\phi$-symmetry is a consequence
of conservation of angular momentum in a central potential.) The
total planetary probability density is thus ``doughnut"
(torus-like) shaped, narrowly peaking around the classical
trajectory.

So, at first sight, it seems like the solar system is perfectly
described as a quantum gravitational system. It even seems
reasonable. Gravity totally dominates as all other forces,
especially the only other known force with infinite reach, the
electromagnetic, cancel due to charge neutrality. The solar system
could thus be seen as a test-vehicle for quantum gravity. In the
solar system the sun totally dominates the gravitational field,
making the central field approximation an excellent one, even
though it in principle is an N-body problem. Contrast this to the
case of multielectron atoms in atomic physics where all electrons
carry the same charge (1/N of the charge of the nucleus), making
the central field approximation a very bad one.

However, from a quantum gravity standpoint, the system could be in
any and all of the degenerate states, and usually at the same
time, so typical for quantum mechanical superposition. Even for
given energy and angular momentum there is no reason for the
planets to be in any particular eigenstate at all of the $2l +1$
allowed, and certainly not exclusively $m = \pm l$. The radial
probability distribution in general has $n-l$ maxima. Thus, only
for $l = l_{max} = n-1$ has it got a unique, highly peaked
maximum. The degeneracy for a given $n$ is $n^2$. Whenever $l <
l_{max}$, the radial wavefunction is highly oscillatory in $r$ as
it has $n-l$ nodes. The same goes for the angular distribution as
there in general are $l-m$ nodes in the $\theta$-direction. For a
general $R_{nl} Y_{lm}$ the planets could be ``all over the
place", and if this weren't bad enough, according to quantum
mechanics the solar system more probable than not should be in
simultaneous, co-existing superposed states with different quantum
numbers as is generic in atomic physics. Consequently, newtonian
quantum gravity cannot solve the quantum mechanical measurement
problem, perhaps because it lacks the non-linear terms conjectured
to be needed \cite{Penrose}.

To get the innermost allowed physical orbit for any
``test-particle", $m$, we must impose the physical restriction
that the binding energy cannot exceed the test particle energy,
thus
\begin{equation}
E_g (max)  = mc^2.
\end{equation}
As $E_g$ can be written
\begin{equation}
E_g = \frac{GmM}{2 b_0},
\end{equation}
we get
\begin{equation}
b_0 (min) = \frac{GM}{2 c^2} = \frac{R_S}{4},
\end{equation}
where $R_S = 2GM/c^2$ is the Schwarzschild radius. The expression
$b_0 (min)$ gives a limit for $b_0$ of the system to be physically
attainable. It is amusing to see how close $b_0 (min)$ is to $R_S$
and one cannot help speculate that a more complete theory of
quantum gravity could ensure that $r
> R_S$ always, and thus forbid black holes altogether\footnote{For the
hydrogen atom the corresponding value is $a_0 (min) \simeq 1.4
\times 10^{-15}$ m, or one-half the ``classical electron radius",
whereas $R_S \simeq 10^{-53}$ m, so that $a_0 (min) \gg R_S$. But
we implicitly already knew that. The Coulomb force does not turn
atoms into black holes.}. The object $M$ must be put together
somehow, but if $r_{min}
> R_S$ it can never accrete enough matter to become a black hole,
as the infalling mass (energy) instead will be radiated away in
its totality (in gravitons), making a black hole state
impossible\footnote{For the classical case, the relation is even
closer, $GmM/r_{min} = mc^2$, giving $r_{min} = R_S /2$, but then
one cannot really speak of energy being carried away by
gravitons.} \cite{Hansson}. This would, in an unexpected way,
resolve the black hole information loss paradox. Even though
$r=R_S$ represents no real singularity, as it can be removed by a
coordinate transformation, anything moving inside $r < R_S$ will,
according to classical general relativity, in a (short) finite
proper time reach the true singularity at $r=0$. If quantum
gravity could ensure that $r > R_S$ always, gravity would of
course be singularity free.

Let us also briefly look at radiative transitions. From the dipole
approximation in atomic physics an elementary quantum (photon)
transition requires $\Delta l = \pm 1$. A quadrupole (graviton)
approximation in quantum gravity instead requires $\Delta l = \pm
2$. So, a typical elementary energy transfer in a highly excited,
gravitationally bound 2-body quantum gravitational system is
\begin{equation}
\Delta E  = - E_g (n^{-2} - (n-2)^{-2}) \simeq \frac{4 E_g}{n^3}.
\end{equation}
For the earth-sun system this means $\Delta E \simeq 2 \times
10^{-20}$ eV, carried by a graviton with frequency $\nu \simeq 5
\times 10^{-6}$ Hz, and wavelength $\lambda \simeq 6 \times
10^{13}$ m $\simeq 400$ AU (1 AU being the mean distance between
the earth and sun).

The average time required for each elementary quantum gravity
transition to take place can be estimated roughly by $\Delta t
\sim \hbar/\Delta E \simeq 3 \times 10^4$ s $\simeq$ 8h 20min.
Thus the power radiated by a spontaneously emitted individual
graviton is very roughly $\sim 10^{-43}$ W, compared to the
prediction from the usual quadrupole formula (first non-vanishing
contribution) in classical general relativity of $\simeq 300$ W
for the total power. We also see that the gravitational force is
not really conservative, even in the static newtonian
approximation, but the difference is exceedingly small in the
sun-earth system. The changes in kinetic and potential energies do
not exactly balance, $\Delta K \neq \Delta U$, the difference
being carried away by gravitons in steps of $\Delta l = 2$. Also,
in quantum gravity there is gravitational radiation even in the
spherically symmetric case, which is forbidden according to the
classical general relativistic description.

Let us now return to the experiment with neutrons in the
gravitational field of the earth \cite{Nezvishevsky}, claiming to
have seen, for the first time, quantum gravitational states in the
potential well formed by the approximately linear gravitational
potential near the earth surface and a horizontal neutron mirror.
An adjustable vertical gap between the mirror and a parallel
neutron absorber above was found to be non-transparent for
traversing neutrons for separations less than $\sim 15 \mu$m
(essentially due to the fact that the neutron ground state
wavefunction then overlaps the absorber). As the neutron in such a
well, from solving the Schr\"{o}dinger equation, has a ground
state wavefunction peaking at $\sim 10 \mu$m, with a corresponding
energy of $\simeq 1.4 \times 10^{-12}$ eV, the experimental result
is interpreted to implicitly having verified, for the first time,
a gravitational quantum state.

If we instead analyze the experiment in the framework of the
present article, the same experimental setup gives $b_0 \simeq 9.5
\times 10^{-30}$ m, $E_g \simeq 2.2 \times 10^{35}$ eV. Close to
the earth's surface, $\tilde{r} \simeq R_{\oplus} \simeq 6.4
\times 10^6$ m, the radius of the earth, giving $n \simeq 8.2
\times 10^{17}$, resulting in a typical energy for an elementary
quantum gravity transition $\Delta E \simeq 4 E_g /n^3 \simeq 1.6
\times 10^{-18}$ eV. For a cavity of $\Delta \tilde{r} = 15 \mu$m,
and $n \gg \Delta n \gg 1$, one gets $\Delta E = E_g b_0 \Delta
\tilde{r} /\tilde{r}^2 \simeq 0.7 \times 10^{-12}$ eV = 0.7 peV,
to be compared to the value 1.4 peV as quoted in
\cite{Nezvishevsky}. Even though the present treatment gives a
similar value for the required energy, it need \textit{not} be the
result of a \textit{single} quantum gravity state as calculated in
\cite{Nezvishevsky}, but rather $\leq 10^6$ gravitons can be
emitted/absorbed. From the treatment in this article it is thus
not self-evident to see why the experimental apparatus
\cite{Nezvishevsky} should be non-transparent to neutrons for
vertical separations $\Delta h < 15 \mu$m.

% LL-, LS-coupling??
%
%TRANSITIONS? (LIFETIMES)? i) Dipole (Rydberg)?, ii) Quadrupole?
%(graviton)
%
%COLLELA-OVERHAUSER-WERNER ("COW") EXPERIMENT???
%
%POUND-REBKA EXPERIMENT???
%
%TITIUS-BODE'S LAW???

To appreciate the potential importance of this, let us digress
briefly on the equivalence principle, the main conceptual pillar
of general relativity. A classic example for illustrating it
involves two rockets: One rocket stands firmly on the surface of
the earth, while the other accelerates constantly in empty space
with $a = g$. According to the equivalence principle there is no
way to, locally, distinguish one from the other if one is not
allowed/able to make outside observations, meaning that
acceleration and gravity are equivalent. However, in the quantum
gravity case, for example considering a neutron inside each
rocket, there certainly \textit{is} a difference: For the rocket
standing on the earth the potential in the Schr\"{o}dinger
equation is $V = -GmM/r$, resulting in normal quantization as
elaborated above. For the rocket accelerating in space, however,
$V = 0$ and the energy levels are non-quantized (the neutron being
a free particle until it hits the floor of the rocket). So the
conclusion is that newtonian quantum gravity breaks the
equivalence principle. Furthermore, to understand why a
free-falling object classically accelerates radially downward in,
e.g., the earth's gravitational field, the gravitons must, by
conservation of momentum, be emitted in the direction opposite to
the acceleration (at least the probability for emission must peak
in that direction). Also, the quantum states with given $n,l,m$
are in principle inherently stable, an outside perturbation being
needed for the transition rate to be different from zero, just
like in atomic physics. For macroscopic bodies this poses no
problem as there in that case are abundant backgrounds of both
gravitational and non-gravitational disturbances. For an
elementary quantum gravity interaction, however, this problem
seems much more severe, as the notion of free-fall loses its
meaning as the quantum states become practically stable to
spontaneous graviton emission. In fact, a bound quantum
gravitational object does not fall at all as it is described by a
stationary wavefunction, or a superposition of such.

Thus, the difference regarding quantized energy levels for an
experiment with neutrons ``falling" under the influence of earth's
gravity \textit{with} mirror (as in \cite{Nezvishevsky}) or
\textit{without} (above) shows that newtonian quantum gravity is
dependent on global boundary conditions, where the boundary in
principle can lie arbitrarily far away. This comes as no surprise,
as the Schr\"{o}dinger equation models the gravitational
interaction as instantaneous, contrasted with the case in general
relativity where the behavior in free-fall only depends on the
local properties of mass-energy and the resulting spacetime
curvature (out of which the mirror is not part due to its
inherently non-gravitational interaction with the neutron) and
causal connection as the gravitational interaction propagates with
the speed of light. However, as several experiments on entangled
quantum states, starting with Clauser/Freedman \cite{Clauser} and
Aspect et al. \cite{Aspect}, seem to be compatible with a
non-local connection between quantum objects \cite{Bell}, this
property of the Schr\"{o}dinger equation does not, at least for
the moment, seem to be a serious drawback for a theory of quantum
gravity. One could even envisage a ``delayed choice" experiment
\'{a} la Wheeler, where the mirror is removed/inserted before the
neutron reaches its position, meaning that we could alter the
energy of the gravitons \textit{after} they have been emitted.

%\begin{figure}[h]
%\begin{center} \psfig{file=fig1.eps}
%\leavevmode
%\epsffile{logistic.eps}
%\end{center}

% \caption{}
% \end{figure}

\end{document}